\documentclass[a4paper,fleqn,usenatbib]{mnras}

\usepackage{newtxtext,newtxmath}
\usepackage{threeparttable}
\usepackage[T1]{fontenc}
\usepackage{ae,aecompl}
\usepackage{threeparttable}

\usepackage{amssymb}	% Extra maths symbols
\usepackage{graphics}
\usepackage{graphicx}	% Including figure files
\usepackage{color,soul}
\usepackage{lscape,rotating}
\usepackage{hyperref}

\usepackage{color}

\usepackage[normalem]{ulem}  % \sout{old text} for strikeout
\usepackage{color,soul}

\def\asec{\ifmmode ^{\prime\prime}\else$^{\prime\prime}$\fi}
\def\msun{M$_{\odot}$}

\def\it{\sl}
\def\degs{\ifmmode ^{\circ}\else$^{\circ}$\fi}
\def\amin{\ifmmode ^{\prime}\else$^{\prime}$\fi}
\def\asec{\ifmmode ^{\prime\prime}\else$^{\prime\prime}$\fi}
\def\fss{\hbox{$.\!\!^{\rm s}$}}        % Fractions of seconds
\def\farcs{\hbox{$.\!\!^{\prime\prime}$}}  % Fractions of arcseconds
\def\h{$^{\rm h}$}
\def\m{$^{\rm m}$}

\def\psra{J0557$+$1550}
\def\psrb{J1630$+$3734}
\def\psrc{J1741$+$1351}
\def\psrd{J2042$+$0246}
\def\gtc{Gran Telescopio Canarias}

\def\fermi{\textit{Fermi}}

\def\degs{\ifmmode ^{\circ}\else$^{\circ}$\fi}
\def\amin{\ifmmode ^{\prime}\else$^{\prime}$\fi}

\def\eqalign#1{\null\,\vcenter{\openup1\jot \m@th
   \ialign{\strut\hfil$\displaystyle{##}$&$\displaystyle{{}##}$\hfil
   \crcr#1\crcr}}\,}

\title[Optical companions to four binary MSPs]{Searching for optical companions to four binary millisecond pulsars with the Gran Telescopio Canarias}

\author[A. Yu. Kirichenko et al.]{A. Yu. Kirichenko$^{1,2}$\thanks{E-mail: aida.taylor@gmail.com},
A. V. Karpova$^{2}$,
D. A. Zyuzin$^{2}$,
S. V. Zharikov$^{1}$,
E. A. L\'opez$^{3}$,\newauthor
Yu. A. Shibanov$^{2}$,
P. C. C. Freire$^4$,
E. Fonseca$^5$,
and A. Cabrera-Lavers$^{6,7}$\\
$^1$Instituto de Astronom\'ia, Universidad Nacional Aut\'onoma de M\'exico, Apdo. Postal 877, Baja California, M\'exico, 22800\\
$^2$Ioffe Institute, Politekhnicheskaya 26, St. Petersburg, 194021,  Russia \\
$^3$Instituto de Investigaci\'on en Ciencias F\'isicas y Matem\'aticas, USAC, Ciudad Universitaria, Zona 12, Guatemala\\
$^4$Max-Planck-Institut f\"ur Radioastronomie, Auf dem H\"ugel 69, D-53131 Bonn, Germany\\
$^5$Department of Physics \& McGill Space Institute, McGill University, 3600 University Street, Montreal, QC, H3A 2T8, Canada\\
$^6$Instituto de Astrof\'isica de Canarias, V\'ia L\'actea s/n, E38200, La Laguna, Tenerife, Spain\\
$^7$GRANTECAN, Cuesta de San Jos\'e s/n, E-38712, Bre\~{n}a Baja, La Palma, Spain
}

\date{Accepted 2020 January 8. Received 2020 January 7; in original form 2019 November 11}

\pubyear{2019}

\begin{document}
\label{firstpage}
\pagerange{\pageref{firstpage}--\pageref{lastpage}}
\maketitle

%%%%%%%%%%%%%%%%%%%%%%%%%%%%%%%%%%%%%%%%%%%%
%%%%%%%%%%%%%%%%%%%%%%%%%%%%%%%%%%%%%%%%%%%%
%The abstract should be a single paragraph not more than 250 words (200 words for Letters).
%No references should appear in the abstract.
\begin{abstract}
We report on multi-band photometric observations of four binary millisecond pulsars with the \gtc. 
The observations led to detection of binary companions to PSRs \psrb, \psrc\ and \psrd\
in the Sloan $g'$, $r'$ and $i'$ bands. Their magnitudes in 
the $r'$ band are $\approx$24.4, 24.4 and 24.0, respectively. 
We also set a 3$\sigma$ upper limit on the brightness of the PSR \psra\ companion in the $r'$ band of $\approx$25.6 mag.
Combining the optical data with the radio timing measurements and  white dwarf cooling models, 
we show that the detected  companions 
are cool low-mass white dwarfs with temperatures and ages in the respective ranges of (4--7)$\times 10^3$~K and 2--5~Gyr. 
All the detected white dwarfs are found to likely have either pure hydrogen or mixed helium-hydrogen atmospheres.     

\end{abstract}


%%%%%%%%%%%%%%%%%%%%%%%%%%%%%%%%%%%%%%%%%%%%%%%%%%

%%%%%%%%%%%%%%%%% BODY OF PAPER %%%%%%%%%%%%%%%%%%
\section{Introduction}

%%%%%%%%%%%%%%%%%%%%%%%%%%%%%%%%%%%%%%%%% Table MSP parameters %%%%%%
\begin{table*}
\caption{Parameters of the binary MSP systems studied in this Paper.
Numbers in parentheses are 1$\sigma$ uncertainties related to the last significant digits quoted.
$D_{\rm YMW16}$ is the DM-distance derived using the YMW16 \citep*{ymw} Galactic electron density model$^a$.
$D_p$ is the timing parallax distance.
} 
\begin{tabular}{lcccc}
\hline
MSP                                                                      & \psra                       & \psrb                         & \psrc                          & \psrd\\
\hline
Right ascension $\alpha$ (J2000)                                         & 05\h57\m31\fss44918(9)      & 16\h30\m36\fss46693(7)      & 17\h41\m31\fss144731(2)        & 20\h42\m11\fss00287(5)     \\
Declination $\delta$ (J2000)                                             & 15\degs50\amin06\farcs046(8) & 37\degs34\amin42\farcs097(1) & 13\degs51\amin44\farcs12188(4) & 02\degs46\amin14\farcs397(2)\\
Galactic longitude $l$ (deg)                                             & 192.68                      & 60.24                         & 37.89                          & 48.99                      \\
Galactic latitude $b$ (deg)                                              & $-$4.31                     & 43.21                         & 21.64                          & $-$23.02                   \\
Proper motion $\mu_\alpha =\dot{\alpha}{\rm cos}\delta$ (mas yr$^{-1}$)  & --                          & 2.4(3.5)                      & $-$8.98(2)                     & 15.1(1.2)                  \\
Proper motion $\mu_\delta$ (mas yr$^{-1}$)                               & --                          & $-$15.9(3.4)                  & $-$7.42(2)                     & $-$14.1(3.5)               \\
\hline
Epoch (MJD)                                                              & 56346                       & 56136                         & 56209                          & 56520                      \\
Spin period $P$ (ms)                                                     & 2.5563670767830(5)          & 3.3181121293936(9)            & 3.747154500259940030(7)        & 4.5337267023282(3)         \\
Period derivative $\dot{P}$ ($10^{-20}$ s s$^{-1}$)                      & 0.735(2)                    & 1.077(9)                      & 3.021648(14)                   & 1.403(6)                   \\
Characteristic age $\tau=P/2\dot{P}$ (Gyr)                               & 5.5                         & 4.9                           & 2.0                            & 5.1                        \\
Orbital period $P_b$ (days)                                              & 4.846550440(4)              & 12.52502574(2)                & 16.335347828(4)                & 77.2005806(3)              \\
Projected semi-major axis $x$ (lt-s)                               & 4.0544597(8)                & 9.039336(1)                  & 11.0033159(5)                  & 24.5699969(8)              \\
Spin-down power $\dot{E}$ (erg s$^{-1}$)                                 & 1.7$\times10^{34}$          & 1.2$\times10^{34}$            & 2.3$\times10^{34}$             &   5.9$\times10^{33}$   \\ 
\hline
Dispersion measure (DM, pc cm$^{-3}$)                                    & 102.6                       & 14.2                          & 24.2                           & 9.3                        \\
Distance $D_{\rm YMW16}$ (kpc)                                             & 1.83                        & 1.19                          & 1.36                         & 0.64                       \\
Timing parallax (mas)                                                    & --                          & --                            & 0.6(1)                         & --                         \\
Distance $D_{p}$ (kpc)                                                   & --                          & --                            & $1.8^{+0.5}_{-0.3}$            & --                         \\
\hline
Companion mass $M_c$ (M$_\odot$)                                         & $\ge 0.2^b$          &           $\ge 0.24^b$                      & $0.22^{+0.05}_{-0.04}$     &    $ \ge 0.19^b$                   \\
Pulsar mass $M_p$ (M$_\odot$)                                            & --                          & --                            & $1.14^{+0.43}_{-0.25}$         & --                         \\
System inclination $i$ (deg)                                             & --                          & --                            & $73^{+3}_{-4}$                 & --                         \\
References$^c$                                                           & [1]                         & [2]                           & [3]                            & [2]                        \\
\hline 
\end{tabular}
\label{tab:param}
\begin{tablenotes}
\item $^a$In the case of the NE2001 model \citep{ne2001}, the distances $D_{\rm NE2001}$ are 2.92 (\psra), 0.94 (\psrb), 0.9 (\psrc) and 0.83 (\psrd) kpc.
\item $^b$A minimum companion mass is calculated assuming the  inclination angle of the binary orbit $i=90$\degs\ and the pulsar mass $M_p=1.4$M$_\odot$. 
\item $^c$Parameters are obtained from [1] -- \citet{scholz2015},
[2] -- \citet{sanpaarsaphd} and
[3] -- \citet{arzoumanian2018} (see also \url{https://data.nanograv.org} for current values).
\end{tablenotes}
\end{table*}

%%%%%%%%%%%%%%%%%%%%%%%%%%%%%%%%%%%%%%%%% Table MSP parameters %%%%%%

Millisecond pulsars (MSPs) represent a subclass of radio pulsars 
that are characterised by short rotational periods ($P<30$ ms) 
and low spin-down rates ($\dot{P}\sim10^{-19}-10^{-20}$ s~s$^{-1}$).
To date, about 400 MSPs have been detected\footnote{\url{https://apatruno.wordpress.com/about/millisecond-pulsar-catalogue/}}.
According to the commonly accepted `recycling' scenario, such objects are formed
as `normal' pulsars in binary systems and then are spun-up due to accretion 
of matter from companion stars \citep{Bisnovatyi-Kogan1974,alpar1982}.
Resulting systems have essentially circular orbits 
and `peeled' companions, usually low-mass white dwarfs (WDs) 
\citep[e.g.][]{tauris2011}. 
To explain  binary MSPs with eccentric orbits, other formation channels are discussed  such as the triple-system formation scenario and the direct formation via a delayed accretion-induced collapse of massive WDs \citep{tauris2011,PortegiesZwar2011,freire2013}. 
Some other possibilities for the origin of these systems are discussed by \citet{2014ApJ...797L..24A}.

Under certain conditions, which are usually defined by the timing noise and MSP binary period, 
the fundamental parameters of MSP systems, including masses and system inclinations, 
can be derived from radio-timing measurements 
of the relativistic Shapiro delay \citep{shapiro1964}.
Otherwise, combination of timing and optical studies is needed.
In case of a WD or a `spider' class companion, 
optical observations are crucial for understanding  
its nature and evolution of the binary system \citep{vankerkwijk2005}. 
Comparison of optical data with WD theoretical cooling models
can provide its parameters, such as chemical composition of WD atmosphere,
mass, temperature, age, and radial velocity (in case of phase-resolved spectroscopic observations).
Combined together with the timing results, this allows one to 
set constraints on the mass of a NS \citep{vankerkwijk2005, 2012MNRAS.423.3316A, antoniadis2013}.
On the other hand, if masses of a binary system components are determined with a high precision 
from the radio timing, optical observations are useful to constrain other parameters. For instance, 
complemented by the radio timing constraints and the theoretical predictions, optical photometric and spectroscopic 
observations allow one to constrain the distance independently of the radio parallax and DM models. 
For all these reasons, recent gamma-ray and radio discoveries of MSP binaries have triggered intensive studies with large 
optical telescopes 
\citep{bassa2016,dai2017,kirichenko2018,beronya2019}.

In this article we present the results of first deep optical observations of four
binary MSPs with the \gtc\ (GTC): PSRs \psra, \psrb, \psrc\ and \psrd. 
The parameters of the systems and their previous studies 
are reviewed in Section~\ref{Targets}. In Section~\ref{sec:data} we provide a description of our
optical observations and data reduction, 
and the results are presented in Section~\ref{sec:results} 
and are discussed in Sections~\ref{sec:results} and \ref{sec:discussion}. 

%%%%%%%%%%%%%%%%%%%%%%%%%%%%%%%%%%%%%%%%%%%%%%%%%%%%%%%%%%%%%%%%%%%%%%%%%%%%%%%%%%%%%%%%%%%%%%%%%%%%%%%%%%
%%%%%%%%%%%%%%%%%%%%%%%%%%%%%%%%%%%%%%%%%%%%%%%%%%%%%%%%%%%%%%%%%%%%%%%%%%%%%%%%%%%%%%%%%%%%%%%%%%%%%%%%%%
%%%%%%%%%%%%%%%%%%%%%%%%%%%%%%%%%%%%%%%%%%%%%%%%%%%%%%%%%%%%%%%%%%%%%%%%%%%%%%%%%%%%%%%%%%%%%%%%%%%%%%%%%%

\section{Targets}
\label{Targets}

The parameters of the four MSP systems are collected 
in Table~\ref{tab:param}. For each of them, the orbital period 
$\gtrsim$ 4 days  
together with the companion mass or its lower limit of $\ge$ 0.19 M$_\odot$ derived from the radio timing 
imply that the companion is likely a WD. We have selected these systems for optical 
observations considering the WD nature of the companions  
and distances $\lesssim2$~kpc. Below we 
shortly describe previous studies of the targets. 

 {\bf PSR \psra}\ was  
 discovered in the PALFA Galactic plane survey 
using the Arecibo radio telescope \citep{scholz2015}. 
The authors did not find any  optical or infra-red 
counterpart to the  pulsar in the SIMBAD 
database and concluded that the apparent visual magnitude of its  companion has to be $>22$.

{\bf PSRs \psrb}\ and {\bf \psrd}\ were discovered in the radio observations  
of \fermi\ unassociated sources \citep{ray2012,sanpaarsaphd}. 
The most recent timing analyses of these pulsars 
were performed by \citet{sanpaarsaphd}.
After inspecting archival data from the Catalina Sky Survey  
with limiting visual magnitudes of 19.5 -- 21.5, 
\citet{salvetti2017} have reported non-detection of the PSR \psrb\ 
companion. 

{\bf PSR \psrc}\ was discovered with the Parkes radio telescope \citep{jacoby2007} 
and the first Shapiro delay measurements for this pulsar were performed by \citet{freire2006}. 
It was also detected in $\gamma$-rays with the \fermi\ telescope 
\citep{espinoza2013}.
Using the 11-yr data set from the North American Nanohertz Observatory for Gravitational Waves (NANOGrav), 
\citet{arzoumanian2018} measured masses of both PSR \psrc\ and its companion
and the system inclination (see Table~\ref{tab:param}).
Our optical detection of the companion with the GTC and its preliminary analysis have been briefly  reported in a conference paper \citep{zyuzin2019jpcs}.

%%%%%%%%%%%%%%%%%%%%%%%%%%%%%%%%%%%%%%%%% Table Observations %%%%%%
\begin{table*}
 \centering
 \caption{Log of the GTC observations. 
} 
 \label{tab:gtc}
 \begin{tabular}{ccccccc}
 \hline
 Filter & Exposure time (s) & Airmass      & Seeing (arcsec) & Zero-point & Date       & MJD   \\
 \hline
  \multicolumn{6}{c}{PSR \psra}\\
 \hline
  $r'$  & 120$\times$12     & 1.06$-$1.10  & 0.7--0.8        & 29.02(1)   & 14-10-2018 & 58405 \\
%  &  &  &  & \\
 \hline
 \multicolumn{6}{c}{PSR \psrb}\\
 \hline
 $g'$   & 200$\times$15     & 1.02$-$1.06  & 0.7$-$0.8       & 28.77(2)   & 06-06-2018 & 58275 \\
 $r'$   & 120$\times$22     & 1.03$-$1.09  & 0.7$-$0.8       & 28.96(1)   & 05-06-2018 & 58274 \\
 $i'$   & 120$\times$23     & 1.09$-$1.21  & 0.7$-$0.8       & 28.59(1)   & 05-06-2018 & 58274 \\  
 \hline
 \multicolumn{6}{c}{PSR \psrc}\\
 \hline
 $g'$   &  200$\times$15    & 1.04$-$1.08  & 0.7$-$1.1       & 28.66(2)   & 06-06-2018 & 58275 \\
 $r'$   &  120$\times$29    & 1.13$-$1.34  & 0.8$-$1.3       & 28.99(1)   & 13-06-2018 & 58282 \\
 $i'$   &  120$\times$23    & 1.13$-$1.29  & 0.7$-$0.9       & 28.55(1)   & 05-06-2018 & 58274 \\
 \hline
 \multicolumn{6}{c}{PSR \psrd}\\
 \hline
 $g'$   & 200$\times$15     & 1.11$-$1.12  & 0.7$-$0.8       & 28.75(2)   & 03-09-2018 & 58364 \\
 $r'$   & 120$\times$23     & 1.12$-$1.18  & 0.6$-$0.7       & 28.98(1)   & 03-09-2018 & 58364 \\
 $i'$   & 120$\times$23     & 1.12$-$1.20  & 0.6$-$0.9       & 28.55(1)   & 03-09-2018 & 58364 \\  
 \hline
  \end{tabular}
\end{table*}
%%%%%%%%%%%%%%%%%%%%%%%%%%%%%%%%%%%%%%%%% Table Observations %%%%%%

%%%%%%%%%%%%%%%%%%%%%%%%%%%%%%%%%%%%%%%%%%%%%%%%%%%%%%%%%%%%%%%%%%%%%%%%%%%%%%%%%%%%%%%%%%%%%%%%%%%%%%%%%%
%%%%%%%%%%%%%%%%%%%%%%%%%%%%%%%%%%%%%%%%%%%%%%%%%%%%%%%%%%%%%%%%%%%%%%%%%%%%%%%%%%%%%%%%%%%%%%%%%%%%%%%%%%
%%%%%%%%%%%%%%%%%%%%%%%%%%%%%%%%%%%%%%%%%%%%%%%%%%%%%%%%%%%%%%%%%%%%%%%%%%%%%%%%%%%%%%%%%%%%%%%%%%%%%%%%%%

\section{Observations and data reduction} 
\label{sec:data}

The fields of PSRs \psra, \psrb, \psrc\ and \psrd\ were observed\footnote{Programmes GTC4-18AMEX and GTC20-18BMEX, PI A. Kirichenko} in 
June, September and October 2018 under clear sky conditions using 
the Optical System for Imaging and low-intermediate Resolution
Integrated Spectroscopy (OSIRIS\footnote{\url{http://www.gtc.iac.es/instruments/osiris/}})  
instrument at the GTC.
OSIRIS consists of two CCDs and provides a field of view (FoV) of 7.8 arcmin~$\times$~7.8 arcmin   
and a pixel scale of 0.254 arcsec. 
To avoid affection by bad pixels, 5 arcsec dithering
between the individual exposures was used in all observing programmes. In addition, 
short 20 s exposures of each pulsar field in the $r'$ band were obtained to avoid saturation of bright stars that were further used for precise astrometry. 
The details of the observations are presented in Table~\ref{tab:gtc}. 

We performed standard data reduction, including bias subtraction and flat-fielding, using the Image Reduction 
and Analysis Facility ({\sc iraf}) package. The cosmic rays were removed from each individual exposure with the L.A.Cosmic 
algorithm \citep{vandokkum}. For each field and filter, the individual images were then aligned to the best quality image and combined. 
The four targets were exposed on CCD2 providing a FoV of 3.9 arcmin~$\times$~7.8 arcmin. All necessary astrometry and photometry calibrations were performed focusing on this part of the detector. 

The astrometric solutions for the four pulsar 
fields were computed 
using the short  
exposures. 
Sets of 16--30 
relatively bright stars from the Gaia DR2 catalogue \citep{gaia1, gaia2} 
detected with the signal-to-noise ratios 
$\ga 20$ were used as the WCS references. Their position uncertainties on the images and catalogue uncertainties 
were  $\la 50$ mas and $\la 1$ mas, respectively.   
OSIRIS  contains geometrical  distortions increasing 
with the distance from the detector aim-point where the targets were exposed. This hampers   the accurate astrometric transformation.     
To minimise distortion effects, the reference stars 
for PSRs \psra, \psrc\ and \psrd\ fields were selected  within 
1 arcmin of the target positions, numbering 
23, 19 and 16, respectively.  For PSR \psrb, which has the largest Galactic  
latitude, there are no sufficient suitable reference objects 
in its immediate vicinity, and we used 30 stars within 4 arcmin 
of its position.   
In addition, to obtain the astrometric fits, we have  
followed the `general' scheme (linear terms plus distortion) described 
in the OSIRIS User Manual\footnote{\url{http://www.gtc.iac.es/instruments/osiris/media/OSIRIS-USER-MANUAL_v3_1.pdf}}. 
We used the {\sc ccmap} routine, which includes the 
frame shift, rotation and scale factor defined as recommended in the manual. 
Formal $rms$ uncertainties of the resulting astrometric fits 
are presented in Table~\ref{tab:red}. They are compatible with the position uncertainties of reference stars on the images. 
Selection of a larger amount of reference  stars as well as choosing 
their different sets  did not change the  solutions significantly. 
The resulting solutions were applied to the combined images.

The photometric referencing was obtained using Sloan standards 
SA 104-428, SA 110-232 and SA 112-805 from \citet{smith2002} 
observed during the same nights as our targets in all respective bands. 
To determine the magnitude zero-points, we used their measured magnitudes 
and the mean OSIRIS atmosphere extinction coefficients 
$k_g'$~= 0.15 $\pm$ 0.02, $k_r'$~= 0.07 $\pm$ 0.01 and $k_i'$~= 0.04 $\pm$ 0.01. 
To verify the obtained calibration, we have compared the magnitudes of several stars in the pulsar fields
against those in the Sloan Digital Sky Survey (SDSS) and Pan-STARRS DR1 catalogues for all bands. In most of the cases, the OSIRIS and the catalogue magnitudes were consistent within uncertainties. The only discrepancy of $\approx$0.1 mag was found for the $g'$-band magnitudes of the PSR \psrc\ field stars, implying a slightly variable transparency during the night of observations, and we have taken it into account in our calculations.
The resulting zero-points are presented in Table~\ref{tab:gtc}.

%%%%%%%%%%%%%%%%%%%%%%%%%%%%%%%%%%%%%%%%% Fig. Observations %%%%%%
\begin{figure*}
\begin{minipage}[h]{1.\linewidth}
\center{
\includegraphics[width=0.48\linewidth,clip]{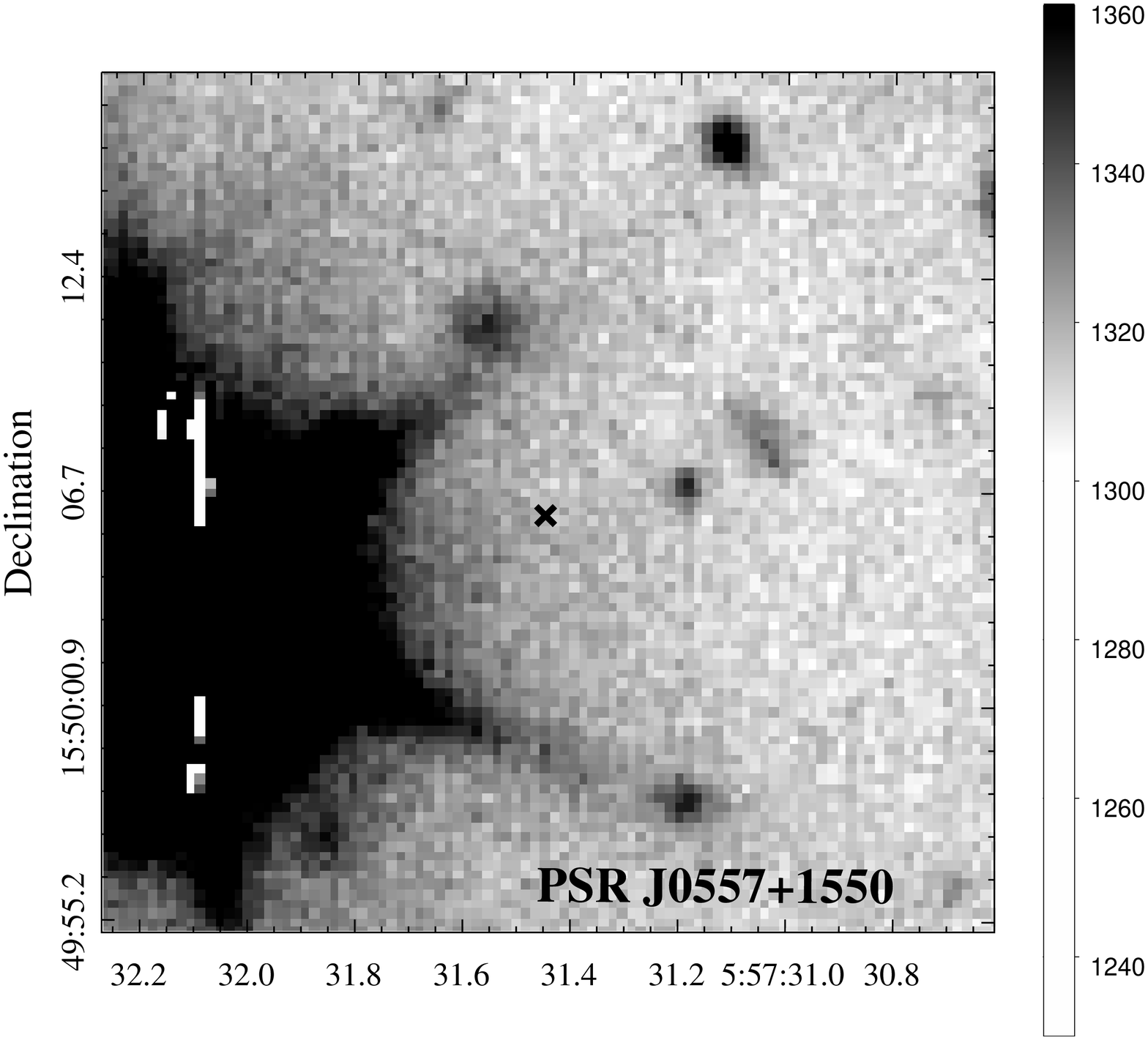}
\includegraphics[width=0.45\linewidth,clip]{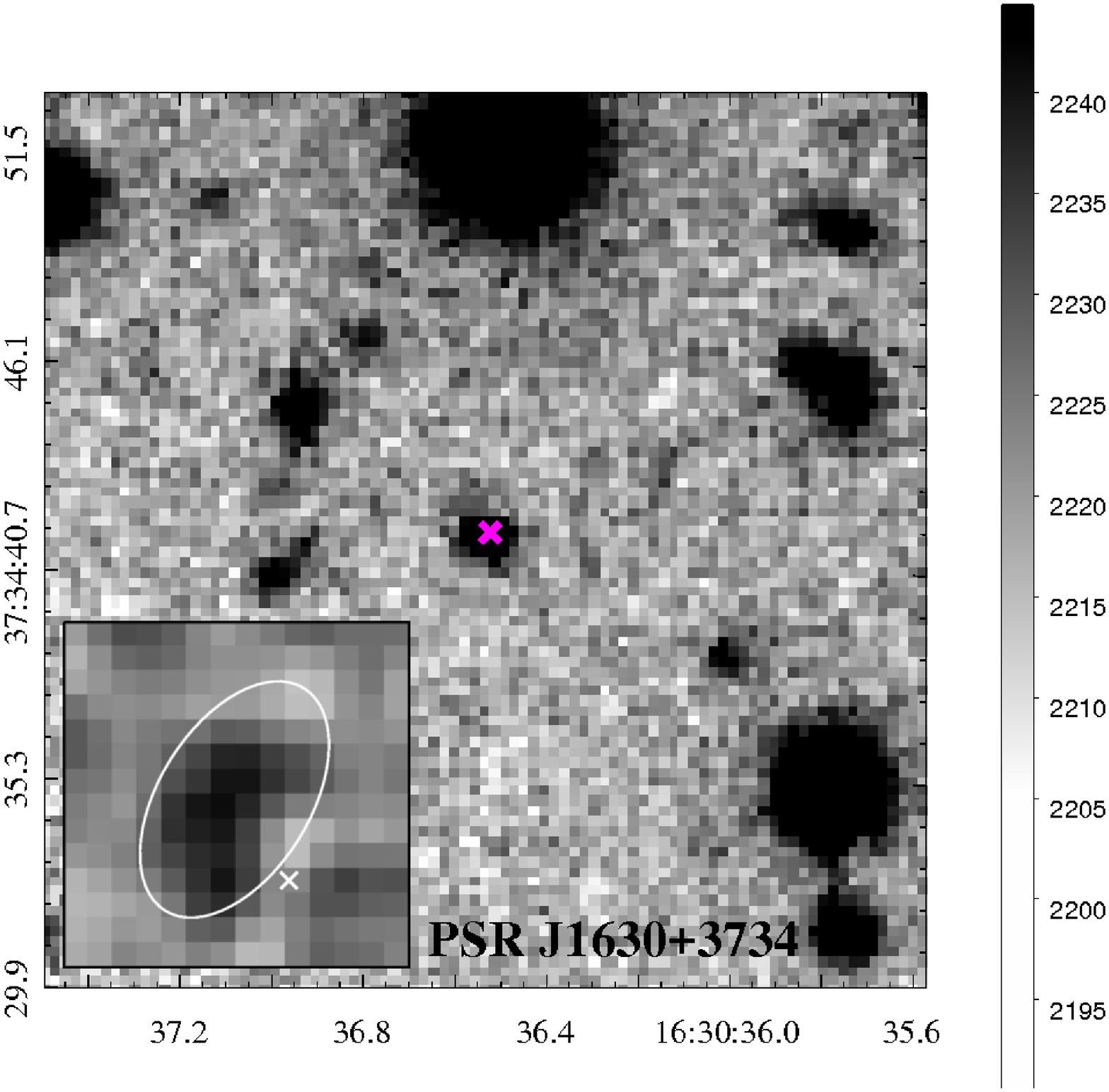}}
\end{minipage}
\begin{minipage}[h]{1.\linewidth}
\center{
\includegraphics[width=0.5\linewidth,clip]{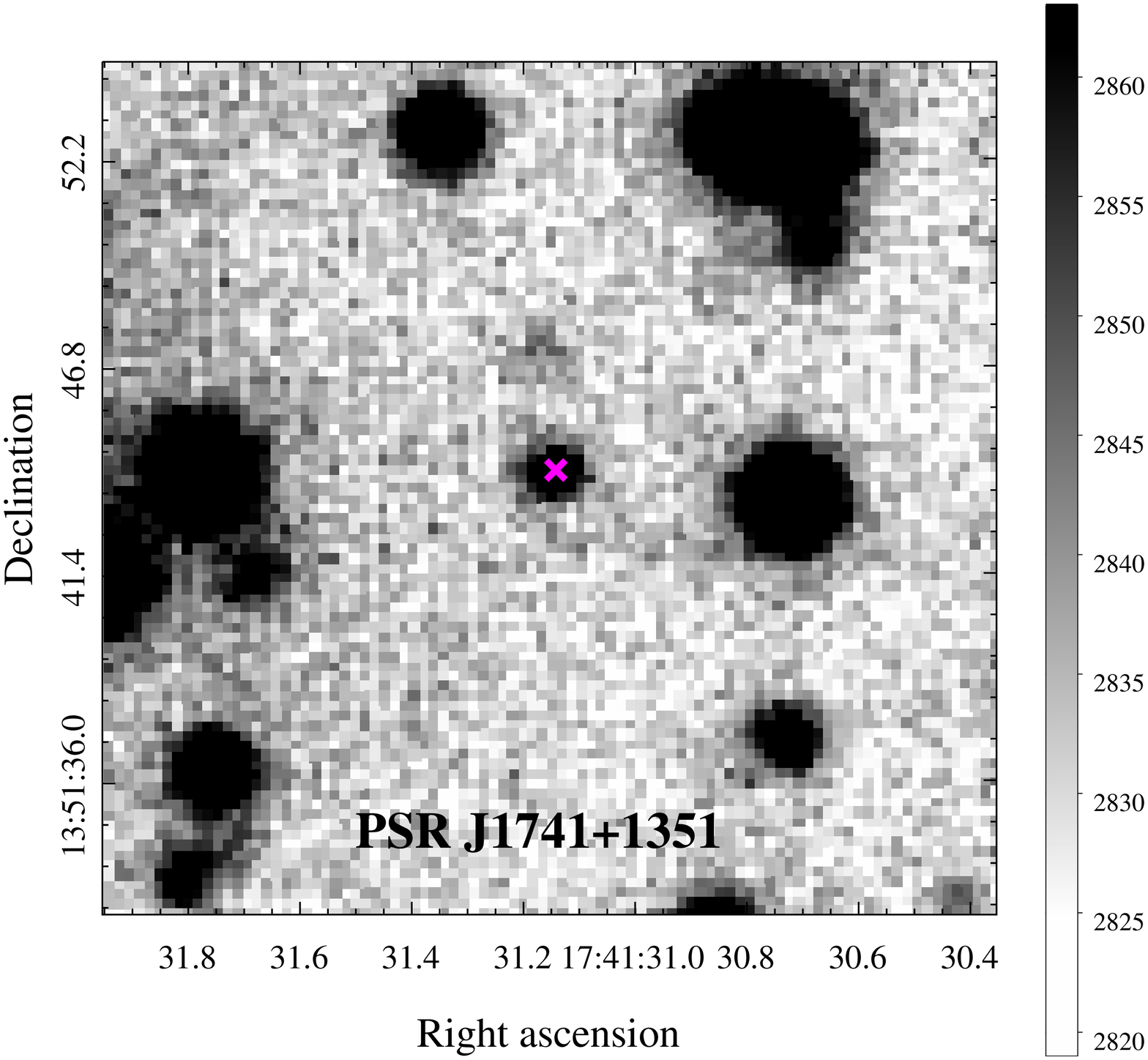}
\includegraphics[width=0.46\linewidth,clip]{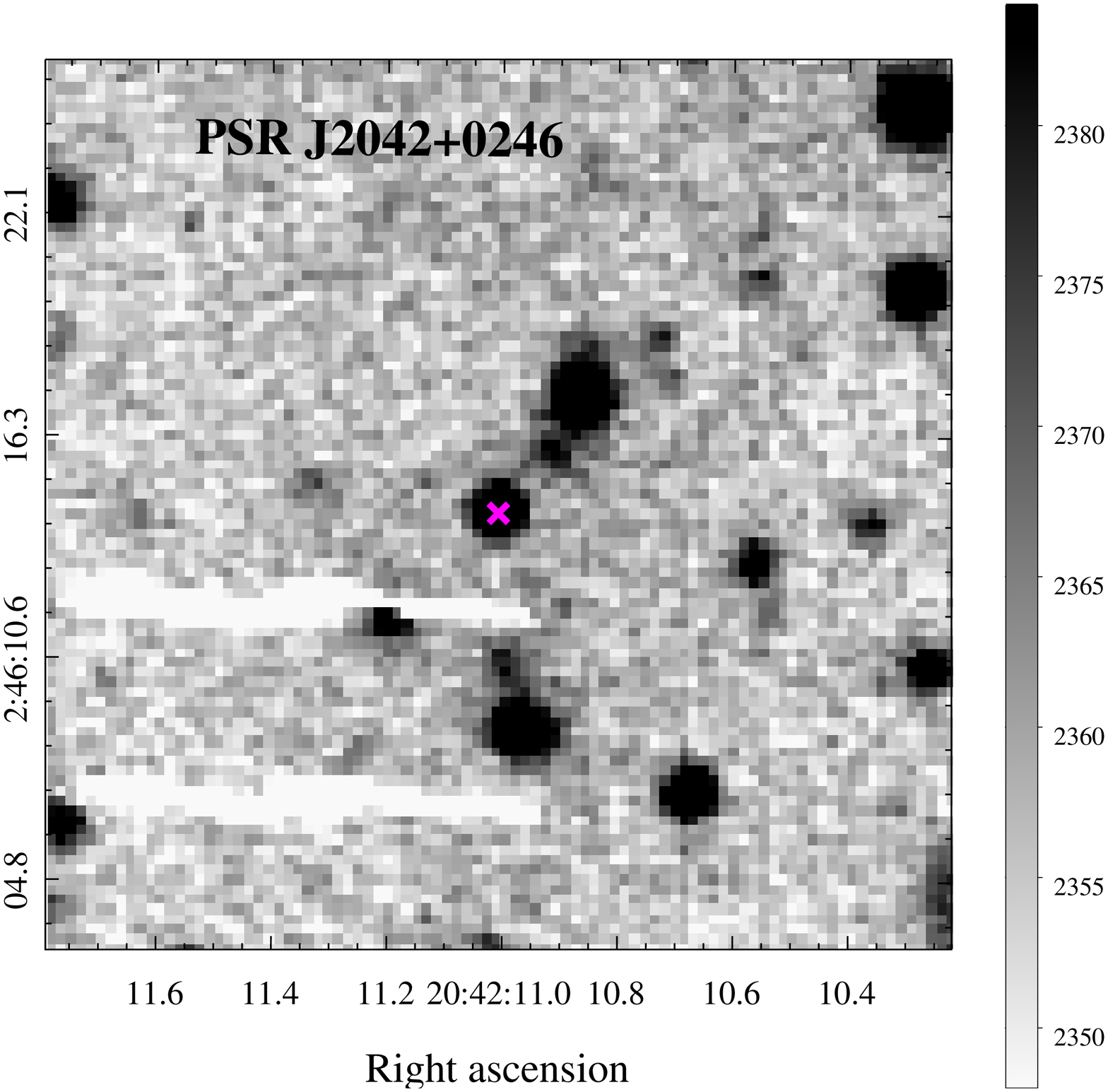}}
\end{minipage}
\caption{GTC/OSIRIS Sloan $r'$-band image fragments of the PSRs \psra, \psrb, \psrc\ and \psrd\ fields. 
The crosses show the pulsar's radio positions for the epoch of the 
optical observations. The PSR \psrb\ 3.5 arcsec~$\times$~3.5 arcsec 
vicinity with the subtracted counterpart is enlarged in the left bottom corner of the respective image fragment. The nearby extended source is enclosed in the ellipse 
and the position of the pulsar is  marked by "X" (see text for details). This fragment  is smoothed with a two-pixel Gaussian kernel. The two white stripes on the PSR \psrd\ image are the detector defects. The astrometric uncertainties of the pulsar positions are negligible in the spatial scale of the images. The colourbar identifies the image values in 100 counts units.}
\label{fig:gtc}
\end{figure*}
%%%%%%%%%%%%%%%%%%%%%%%%%%%%%%%%%%%%%%%%% Fig. Observations %%%%%%

%%%%%%%%%%%%%%%%%%%%%%%%%%%%%%%%%%%%%%%%% Table Reddening %%%%%%
\begin{table*}
\centering
\caption{Astrometric and photometric information for the MSP systems and the interstellar reddening in their direction. 
Reddening, intrinsic colours and absolute magnitudes are provided for the latest published distance estimations, i.e. DM distances for PSRs \psra, \psrb\ and \psrd\ based on the YMW16 model
and the timing parallax distance for PSR \psrc\ (see Table~\ref{tab:param}). 
$\alpha_p$ and $\delta_p$ are the pulsars' coordinates for the epoch of the GTC observations
(errors include uncertainties of the radio positions and proper motion which are negligible in comparison with the \textit{rms}
uncertainties of the GTC astrometric fits).
$\alpha_c$, $\Delta\alpha_c$  and $\delta_c$, $\Delta\delta_c$ are coordinates of the optical companions on the images and their \textit{rms} uncertainties.
}
\label{tab:red}
\begin{tabular}{lcccc}
\hline
MSP                                  & \psra                        & \psrb                       & \psrc                          & \psrd \\
\hline
$\alpha_p$ (J2000)                   & 05\h57\m31\fss44918(9)       & 16\h30\m36\fss468(2)        & 17\h41\m31\fss141245(8)        & 20\h42\m11\fss0080(4)\\
$\delta_p$ (J2000)                   & 15\degs50\amin06\farcs046(8) & 37\degs34\amin42\farcs00(2) & 13\degs51\amin44\farcs0799(1) & 02\degs46\amin14\farcs33(2)\\
\hline
$\alpha_c$ (J2000)                   & --                           & 16\h30\m36\fss475        & 17\h41\m31\fss144              & 20\h42\m11\fss005  \\
$\delta_c$ (J2000)                   & --                           & 37\degs34\amin41\farcs92    & 13\degs51\amin43\farcs95       & 02\degs46\amin14\farcs46 \\ 
\textit{rms} $\Delta\alpha_c$, arcsec  & 0.05                         & 0.02                       & 0.05                           & 0.04           \\
\textit{rms} $\Delta\delta_c$, arcsec  & 0.04                         & 0.02                        & 0.05                           & 0.05           \\
\hline
$g'$                                 & --                           &25.43(4)                     & 24.74(5)                       & 24.96(4)       \\
$r'$                                 & $>$25.6                      & 24.44(4)                    & 24.38(4)                       & 23.97(2)       \\
$i'$                                 & --                           & 24.17(4)                    & 24.17(4)                       & 23.58(3)       \\
\hline            
$E(B-V)^a$                           & $0.19^{+0.03}_{-0.01}$       & $0.00^{+0.01}_{-0.00}$      & $0.12^{+0.01}_{-0.03}$         & $0.07\pm0.02$  \\
$A_{g'}$                             & --                           & $0.00^{+0.03}_{-0.00}$      & $0.42^{+0.03}_{-0.10}$         & $0.26\pm0.07$  \\
$A_{r'}$                             & $0.48^{+0.07}_{-0.02}$       & $0.00^{+0.02}_{-0.00}$      & $0.30^{+0.02}_{-0.07}$         & $0.18\pm0.05$  \\ 
$A_{i'}$                             & --                           & $0.00^{+0.02}_{-0.00}$      & $0.22^{+0.02}_{-0.05}$         & $0.14\pm0.03$  \\
\hline
$(g'-r')_0$                          & --                           & $0.99^{+0.07}_{-0.06}$      & $0.24^{+0.07}_{-0.14}$         & $0.93\pm0.10$  \\
$(r'-i')_0$                          & --                           & $0.27^{+0.06}_{-0.06}$      & $0.13^{+0.06}_{-0.10}$         & $0.35\pm0.07$  \\
$M_{r'}$                             & $>$13.8                      & $14.06\pm0.04$              & $12.80^{+0.44}_{-0.61}$        & $14.76\pm0.05$ \\
\hline
 Effective temperature $T_{\rm eff}$, K & $\lesssim$ 5.0 $\times$ $10^3$ & $\sim 4 \times 10^3$ & $\sim$ (6--7) $\times 10^3$ & 4.49(14)~$\times 10^3$  \\
Cooling age $t$, Gyr  & $\gtrsim$ 1.5 &  $\sim$ 2--5 & $\sim$ 1--2 & 5.6$^{+0.9}_{-1.2}$ \\
\hline
\end{tabular}
\begin{tablenotes}
\item $^a$In the directions to PSRs \psrb\ and \psrd\ the reddening is the same for the DM distance estimations provided either by the YMW16 or NE2001 models.
For \psrc\ $E(B-V)$ does not vary for $D>0.9$ kpc.
\end{tablenotes}
\end{table*}

%%%%%%%%%%%%%%%%%%%%%%%%%%%%%%%%%%%%%%%%%%%%%%%%%%%%%%%%%%%%%%%%%%%%%%%%%%%%%%%%%%%%%%%%%%%%%%%%%%%%%%%%%%
%%%%%%%%%%%%%%%%%%%%%%%%%%%%%%%%%%%%%%%%%%%%%%%%%%%%%%%%%%%%%%%%%%%%%%%%%%%%%%%%%%%%%%%%%%%%%%%%%%%%%%%%%%
%%%%%%%%%%%%%%%%%%%%%%%%%%%%%%%%%%%%%%%%%%%%%%%%%%%%%%%%%%%%%%%%%%%%%%%%%%%%%%%%%%%%%%%%%%%%%%%%%%%%%%%%%%
\section{Results and discussion}
\label{sec:results}

The resulting $\sim$23 arcsec~$\times$~23 arcsec $r'$-band 
images 
of the four pulsar fields are presented in Figure~\ref{fig:gtc}. 
The crosses correspond to the pulsar positions
obtained from the radio timing and shifted according to proper motions to match the epoch of the optical observations, 
where respective information on the proper motion is provided. 
The astrometric uncertainties
are negligible in the spatial scale of the optical images. 

In the fields of PSRs \psrb, \psrc\ and \psrd\ in all bands we firmly detect star-like objects, 
whose coordinates coincide with the pulsar positions (see also Table~\ref{tab:red}). 
To calculate the probability 
of an accidental coincidence of pulsar positional regions with unrelated field objects, 
we have used the Poisson distribution $P$ = 1$-\exp(-\pi \sigma R^2)$, where $\sigma$ and $R$ correspond to the surface number density of stars with a similar magnitude and the astrometric accuracy, respectively.
Considering magnitudes in a range of 19$-$25, in all cases this probability does not exceed $\approx$ $10^{-2}$.
For this reason, below we consider the detected objects as optical binary companions of the MSPs.  
Two of them, companions to PSRs \psrc\ and \psrd, show star-like profiles, while the PSR \psrb\ companion profile is more complex. The point spread function (PSF) subtraction of this   
source using the {\sc iraf/daophot}~{\sc allstar} routine has revealed a faint slightly extended underlying object located about one arcsec north-east 
from the source maximum peak. It is enlarged in the left bottom corner of the right upper panel of Figure~\ref{fig:gtc}, where the companion was subtracted. The PSF photometry and iterative PSF subtruction were applied both to the companion   
and this source to measure their magnitudes and positions. 
The underlying object is detected at about 3$\sigma$ significance in the $r'$ and $i'$ 
bands with a magnitude of $\approx$26.5 and is not detected in the $g'$ band down 
to the $\approx$27.5 magnitude limit. 
Its relation to the pulsar companion remains unclear. 
It is likely that the object is 
 a hint of a distant galaxy. 
The resulting observed magnitudes of the proposed companions are presented in Table~\ref{tab:red}. 

As the PSR \psra\ companion is undetected in the only available $r'$ band,
we have estimated a $3\sigma$ upper limit on its brightness in this band (see Table~\ref{tab:red}). 

To calculate the absolute magnitudes and intrinsic colours of the optical companions,
we used the 3D map of interstellar dust reddening $E(B-V)$ which is based on Pan-STARRS~1 
and 2MASS stellar photometry and Gaia parallaxes \citep{dustmap2019}\footnote{\url{http://argonaut.skymaps.info/}}.
Using distances to the pulsars from Table~\ref{tab:param}, we obtained the corresponding $E(B-V)$ colour excesses (Table~\ref{tab:red}). 
They were then converted to the extinction corrections $A_{g',r',i'}$ for all the bands utilising coefficients
provided by \citet{schlafly2011}. 
The resulting corrections, absolute magnitudes and intrinsic colours are presented in Table~\ref{tab:red}.

\begin{figure*}
\begin{minipage}[h]{0.77\linewidth}
%\center{\includegraphics[width=1.0\linewidth,clip]{col-mag281019.eps}}
\center{\includegraphics[width=1.0\linewidth,clip]{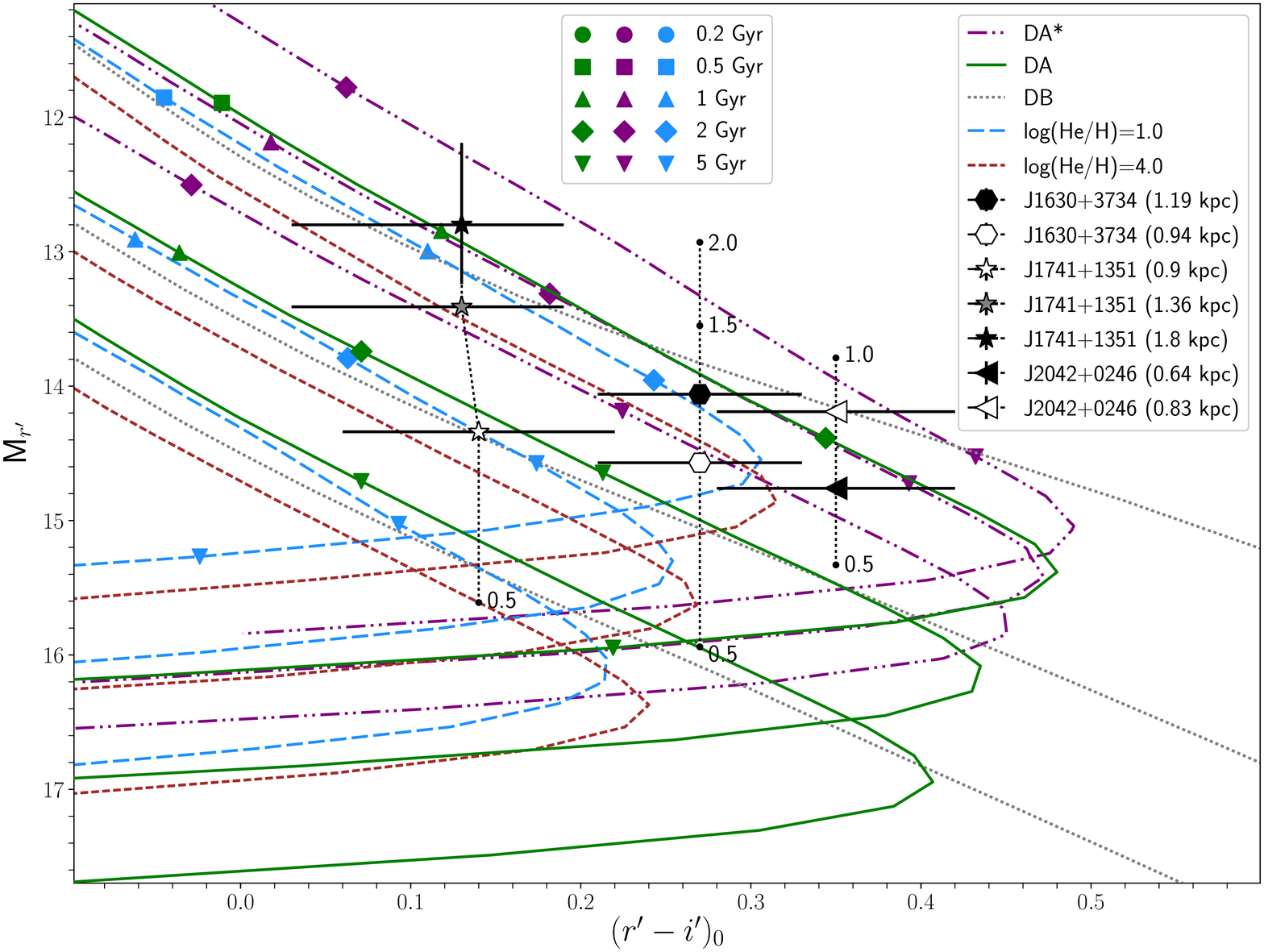}}
\end{minipage}

\begin{minipage}[h]{0.77\linewidth}
%\center{\includegraphics[width=1.0\linewidth,clip]{col-col281019.eps}}
\center{\includegraphics[width=1.0\linewidth,clip]{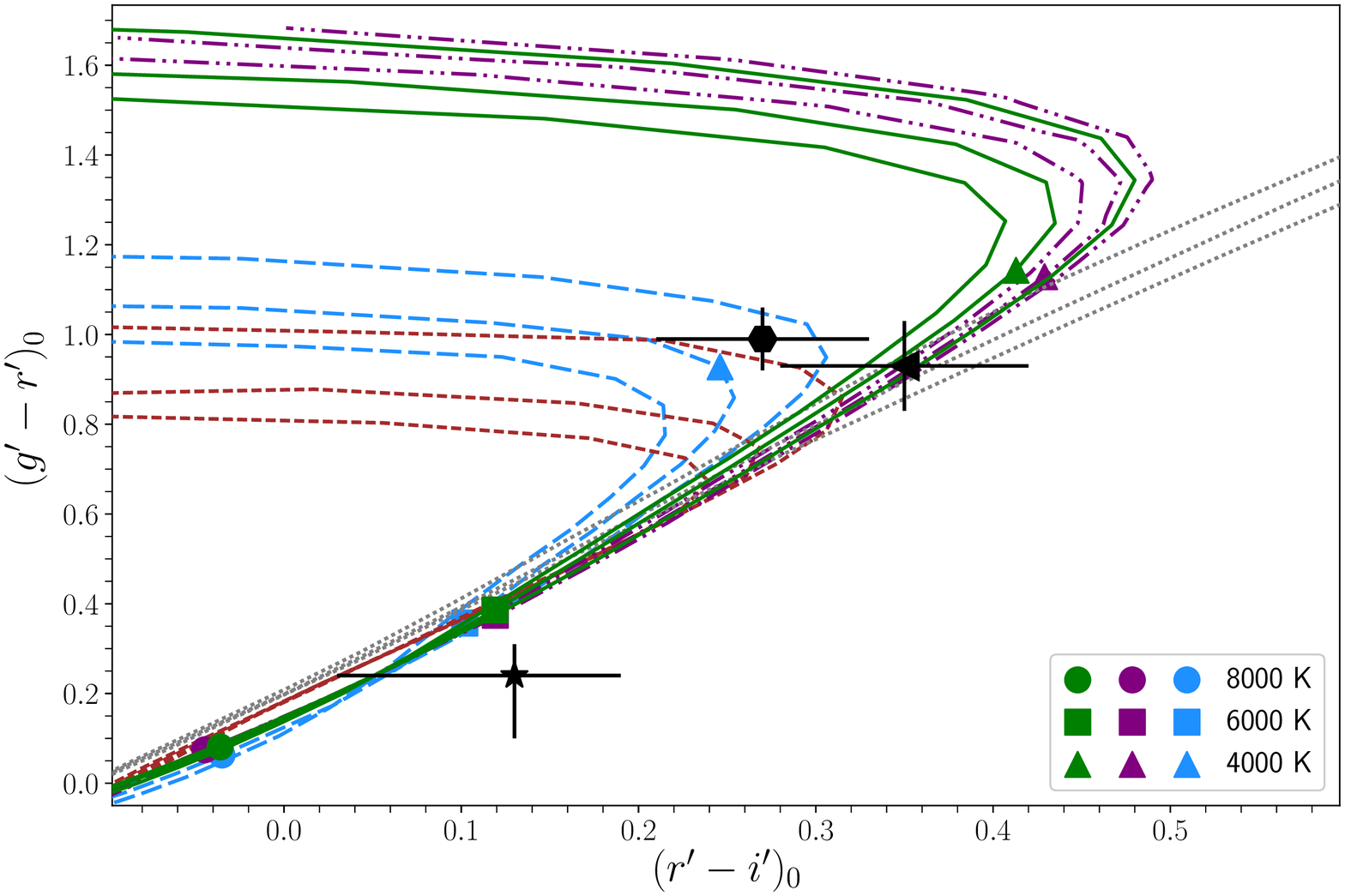}}
\end{minipage}
\caption{Colour-magnitude (top) and colour-colour (bottom) diagrams with various WD cooling tracks 
and data for WD companions to different MSPs listed in the legend. 
Purple dash-dot-dotted lines (DA*) show evolutionary tracks by \citet{althaus2013} 
for helium-core WDs with masses 0.1821, 0.2724 and 0.4352\msun.
Solid green (DA) and %black 
grey dotted (DB) lines show models 
for CO-core WDs with hydrogen and helium atmospheres, respectively,  
\citep{holberg2006,kowalski2006,tremblay2011,bergeron2011} with masses 0.2, 0.6 and 1.0\msun. 
Dashed blue and brown lines are tracks for WDs with mixed atmospheres 
(log(He/H) = 1.0 and 4.0, respectively) and masses 0.2, 0.6 and 1.0\msun.
Masses increase from upper to lower tracks.
WDs' ages (top) and effective temperatures (bottom) are shown by different symbols.
The positions of the PSRs' \psrb, \psrc\ and \psrd\ companions are shown 
by different symbols as indicated in the legend.
In the top panel, for each source its positions are shown for different DM distance estimates
(provided by the YMW16 and NE2001 models); for the PSR \psrc\ companion the position corresponding to 
the timing parallax distance is also indicated (see Table~\ref{tab:param}).
We also indicate their positions at various additional distances 
derived using reddening values from the dust map by \citet{dustmap2019}.
The distances are marked by the numbers in kpc units.
Since their colours do not change significantly with the distance, in the lower panel
the results are presented for the distances from Table~\ref{tab:param}.}
\label{fig:col-mag}
\end{figure*}

As we noted in Sect.~\ref{Targets},  
the companions to the pulsars are likely 
WDs. To verify this, we compared the 
obtained absolute magnitudes and dereddened colours with the cooling models of helium-core 
WDs with hydrogen-atmospheres \citep{althaus2013}\footnote{\url{http://evolgroup.fcaglp.unlp.edu.ar/TRACKS/tracks_heliumcore.html}} 
and CO-core WDs with hydrogen and helium atmospheres
\citep[known as the Bergeron models;][]{holberg2006,kowalski2006,tremblay2011,
bergeron2011}\footnote{\url{http://www.astro.umontreal.ca/~bergeron/CoolingModels}.
}.
The corresponding colour-magnitude and colour-colour diagrams 
are presented in Figure~\ref{fig:col-mag}, where evolutionary sequences 
for WDs of different classes are shown by different line types. 
The positions of the detected companions in the diagrams, particularly in the colour-magnitude diagram,    
depend on the accepted distances to the pulsars. To account for different possibilities, for each pulsar we show 
at least two positions, corresponding  
to the DM-distances based on the  
NE2001 and YMW16 Galactic electron density distribution models (see Table~\ref{tab:param}). 
For PSR \psrc, we also include the point corresponding to 
the parallactic distance from \citet{arzoumanian2018}.        
Bellow we analyse the results obtained for each object.

%%%%%%%%%%%%%%%%%%%%%%%%%%%%%%%%%%%%%%%%% Fig. Col-mag %%%%%%
%%%%%%%%%%%%%%%%%%%%%%%%%%%%%%%%%%%%%%%%% Fig. Col-mag %%%%%%

%%%%%%%%%%%%%%%%%%%%%%%%%%%%%%%%%%%%%%%%% Fig. Col-mag J1741 %%%%%%
\begin{figure}
\begin{minipage}[h]{1.\linewidth}
\center{\includegraphics[width=1.0\linewidth,clip]{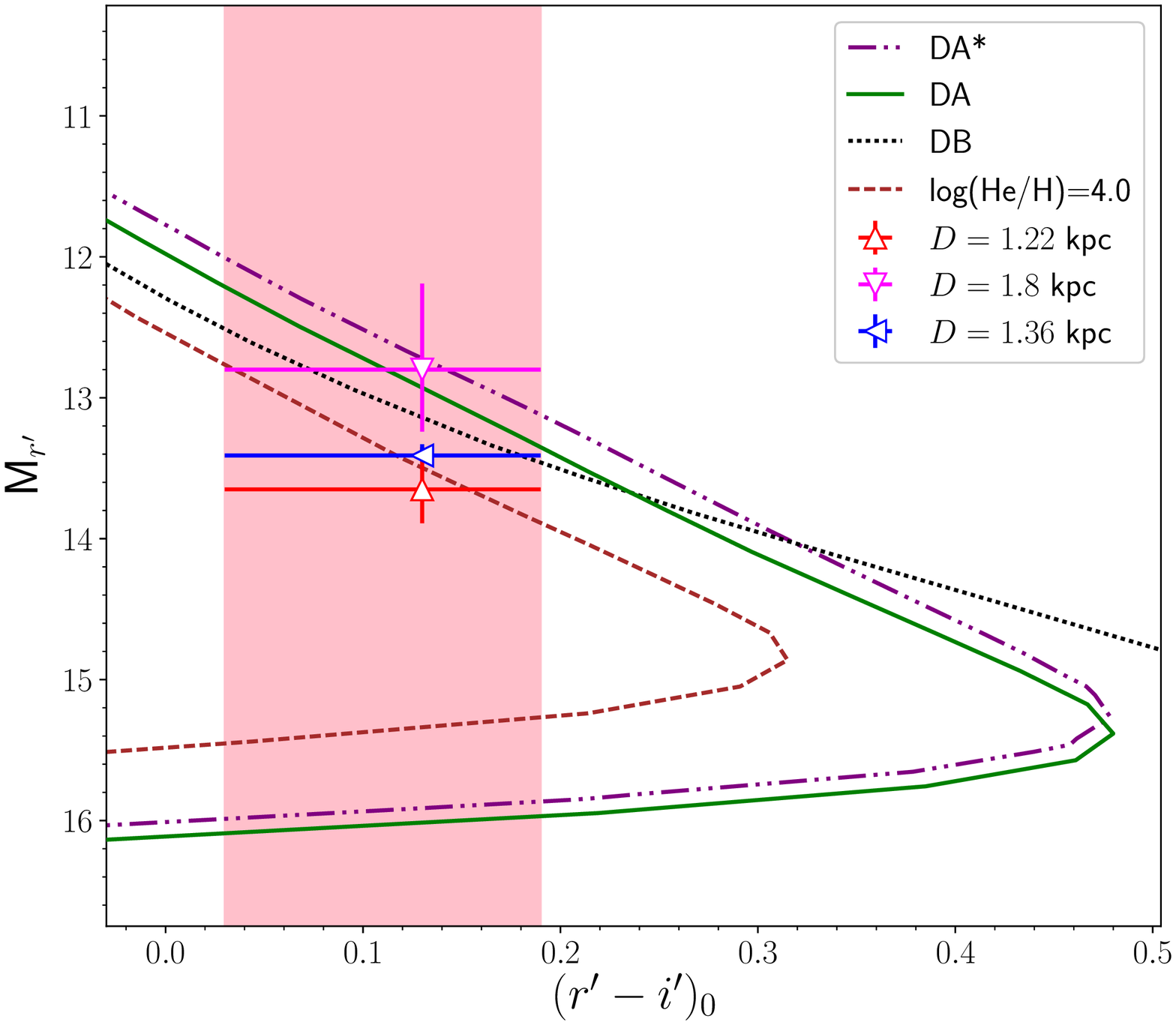}}
\end{minipage}

\caption{Colour-magnitude diagrams with WD cooling sequences. 
The dash-dot-dotted purple line (DA*) shows the track for a helium-core WD with a mass of 0.239\msun\ \citep{althaus2013},
solid green (DA) and black dotted (DB) lines -- 
for CO-core WDs with masses of 0.2\msun\ with hydrogen and helium atmospheres, respectively 
\citep{holberg2006,kowalski2006,tremblay2011,bergeron2011}, 
the dashed brown line -- for a WD with mixed atmosphere 
(log(He/H) = 4.0) and a mass of 0.2\msun.
The positions of the PSR \psrc\ optical companion are shown 
by triangle symbols for different distance estimates indicated in the legend.
The pink stripe corresponds to the source $(r'-i')_0$ colour in the case
of the maximum extinction in the pulsar direction.}
\label{fig:col-mag-1741}
\end{figure}
%%%%%%%%%%%%%%%%%%%%%%%%%%%%%%%%%%%%%%%%% Fig. Col-mag J1741 %%%%%%

%%%%%%%%%%%%%%%%%%%%%%%%%%%%%%%%%%%%%%%%% Fig. Mass & temperature J2042 %%%%%%
\begin{figure}
\begin{minipage}[h]{1.0\linewidth}
\center{\includegraphics[width=1.0\linewidth,clip]{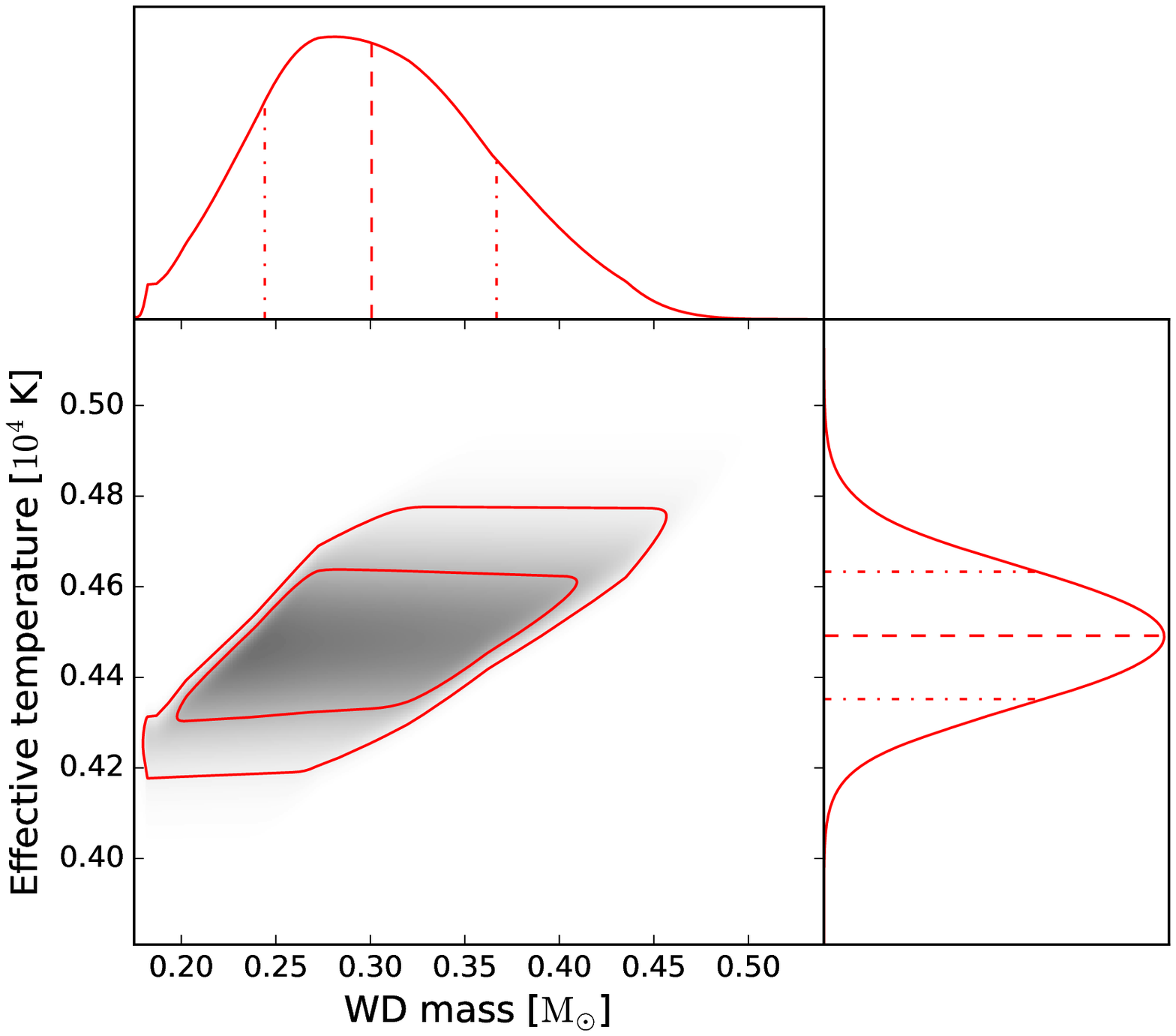}}
\end{minipage}

\caption{Constraints on the effective temperature and mass for the PSR \psrd\ WD companion. 
The contours show the 68.3\% and 95.5\% confidence levels. 
1D likelihoods are shown in the top and side panels. 
Dashed lines indicate the median values and dash-dotted lines are the 1$\sigma$ confidence intervals.
}
\label{fig:mass-temp}
\end{figure}
%%%%%%%%%%%%%%%%%%%%%%%%%%%%%%%%%%%%%%%%% Fig. Mass & temperature J2042 %%%%%%

\subsection{PSR \psra}

PSR \psra\ was observed only in the $r'$ band 
and no optical source was detected at its radio position down to $r'\geqslant 25.6$ mag.
Comparison of the lower limit on $M_{r'}$ (see Table~\ref{tab:red}, $D_{\rm YMW16}=1.83$~kpc) 
with the cooling models for a hydrogen-atmosphere WD with the minimum companion mass $\approx0.2$M$_\odot$ 
implies the age $t\gtrsim$ 1.5--3 Gyr  
and the WD effective temperature $T_{\rm eff}\lesssim (4.6-5.0)\times 10^3$~K.
The respective values for the distance $D_{\rm NE2001}=2.92$ kpc, which is based on the NE2001 model, are 
$M_{r'}>12.6$, $t\gtrsim$1--2 Gyr
and $T_{\rm eff}\lesssim (5.7-6.3)\times 10^3$~K.

\subsection{PSR \psrb}

In the colour-colour diagram, the PSR \psrb\ companion  shows a $\approx2\sigma$ error displacement to bluer colour indices from 
the tracks of WDs with pure hydrogen or helium atmospheres (DA*, DA and DB tracks; see the bottom panel of Figure~\ref{fig:col-mag}).
The shift is not related to uncertainties of the DM distance   
as the extinction in the PSR~\psrb\ direction is very low (see Table~\ref{tab:red}) 
and does not affect the source intrinsic colours. 

It is known that models of ultra-cool WDs with mixed He/H atmospheres show bluer $(g'-r')_0$ 
colours as opposed to those of WDs  with pure hydrogen or helium atmospheres \citep[e.g.][]{parsons2012}.
As an example, in Figure~\ref{fig:col-mag} we also present the cooling tracks 
for CO-core WDs with mixed atmospheres (for log(He/H) = 1.0 and 4.0). 
The position of the PSR \psrb\ companion is in accord with these tracks. 
Unfortunately, respective tracks for helium-core WDs are not available and at the current stage 
the determination of the presumed He/H ratio in the WD atmosphere is not possible.  
If the source is indeed a WD with a mixed atmosphere, 
its temperature and age are $\sim 4\times 10^3$~K and $\sim2$--5 Gyr, respectively. 
We note, however, that the $\sim$2$\sigma$ displacement to bluer colour indices does not exclude the pure hydrogen or helium atmospheres, but makes them less plausible.

\subsection{PSR \psrc}

The temperature of the  PSR \psrc\ companion, assuming its WD nature and taking into account the intrinsic colour indices (see  Figure~\ref{fig:col-mag}), is about (6--7)$\times10^3$ K. This estimation 
is independent of any considered distance from Table~\ref{tab:param} as $E(B-V)$ along the pulsar line of sight and 
the intrinsic colours of the companion vary only slightly within the expected distance range $\approx$ 0.9--2.3~kpc.
The radio timing analysis by \citet{arzoumanian2018} provides the parallax distance to the pulsar $1.8^{+0.5}_{-0.3}$ kpc and the companion mass $M_c=0.22(5)$ \msun.
The corresponding derived absolute magnitude and colour index of the companion are M$_{r'}$ = $12.80^{+0.44}_{-0.61}$ 
and $(r'-i')_0$ = $0.13^{+0.06}_{-0.10}$. 
They are consistent with a hydrogen or helium atmosphere WD with a mass of 0.2--0.3 \msun which is in agreement with the mass measurement from the radio timing.  
The respective WD cooling age is 1--2 Gyr (see Figure~\ref{fig:col-mag}).   
The DM distance $D_{\rm YMW16}=1.36$~kpc based on the YMW16 model implies a helium or mixed He/H atmosphere WD with similar mass and age ranges, 0.2--0.5 \msun\ and 1--2 Gyr, respectively. 
In contrast, the smaller DM distance $D_{\rm NE2001}=0.9$ kpc requires an older ($\sim$2--5 Gyr) and

Recently, Freire et al. (in preparation) have reported measurements of 
the PSR \psrc\ system parameters based on new radio observations with 
high cadence in combination with publicly available data provided by NANOGrav
\citep{2013ApJ...762...94D}.
They confirmed and improved the value for the companion mass $M_c=0.227^{+0.014}_{-0.013}$~\msun, provided the 
estimations on the pulsar mass $M_p=1.19^{+0.11}_{-0.10}$~\msun\
and the parallactic distance $1.22^{+0.13}_{-0.11}$~kpc, however these are still preliminary.   
The new parallactic  distance 
is marginally consistent with the DM distance $D_{\rm YMW16}=1.36$ kpc 
and implies that the PSR~\psrc\ companion may have a mixed atmosphere. 
In Figure~\ref{fig:col-mag-1741} we present a colour-magnitude diagram with selected evolution tracks for WDs with different atmosphere compositions (DA, DB, and He/H) and masses that are close to the companion mass. 
At the distance D = 1.22 kpc, the companion perfectly agrees  
with a WD with a mixed He/H atmosphere. 

Nevertheless, taking into account the formal distance and photometric measurements' uncertainty,
a WD with a pure hydrogen or helium atmosphere is rather less plausible than completely rejected.
As we mentioned before, $E(B-V)$ in the pulsar direction does not vary for the distances $\gtrsim0.9$ kpc \citep{dustmap2019} and the companion colours do not change with the distance either.
The stripe in Figure~\ref{fig:col-mag-1741} corresponds to the source intrinsic colour in case of the maximum $E(B-V)$. 
Pure hydrogen atmospheres of WDs provide higher luminosities and 
larger corresponding distances in a range of 1.5--2.7 kpc, whereas the mixed atmospheres require a distance between 1.1 and 1.52 kpc.
Therefore, determining the distance to the system is critical to select the appropriate model of the companion atmosphere. 

The DA and DB tracks in Figures \ref{fig:col-mag} and \ref{fig:col-mag-1741} represent the CO-core WDs cooling models,  
while it is generally believed that low-mass WDs in MSP binaries have helium cores \citep{tauris2011}. 
Comparison of tracks for (DA$^*$) helium-core and CO-core WDs with hydrogen atmospheres shows 
that at a given age and mass the latter ones are less luminous \citep{vanoirschot2014}. 
At ages $\ga$ 1 Gyr for low-mass WDs the brightness difference becomes less than a half 
of a magnitude. As seen from Figure~\ref{fig:col-mag-1741}, for the specific masses
the difference between the DA and DA* tracks is even smaller and is comparable to the derived uncertainty in the absolute magnitude of the companion.

\subsection{PSR \psrd}

As it was mentioned before, most of the WD companions to MSPs have a pure hydrogen atmosphere. Positions of the PSR \psrd\ companion on the colour-magnitude and colour-colour 
diagrams  in 
Figure~\ref{fig:col-mag}  
roughly correspond to a cool ($\approx$4500~K)  hydrogen-atmosphere WD 
with an age of $\sim$2--5 Gyr, depending on the cooling model.

To better constrain the WD parameters for this system,  we utilised the procedure described by \citet{dai2017} and  
used the models for helium-core hydrogen-atmosphere WDs with masses of  
0.1554--0.4352~\msun\ 
and the 
CO-core hydrogen-atmosphere WDs with masses of 0.5--1.0~\msun. 
The models were interpolated in the mass--temperature plane within a mass range 
of 0.1554--1.0~\msun\ and a temperature range of 3000--10000 K using a 7000$\times$7000 grid.  
Then the likelihood of each model point was calculated 
according to formula (5) from \citet{dai2017}.
We assumed the distance range between $D_{\rm YMW16}$ and $D_{\rm NE2001}$, i.e. 0.6--0.8 kpc.
As a result, 
we derived a WD mass of $0.30^{+0.07}_{-0.06}$~\msun, a temperature of 4.49(14)~$\times10^3$~K and an age of $5.6^{+0.9}_{-1.2}$ Gyr (the values correspond to the  
medians of the probability distributions and their 1$\sigma$ uncertainties).
The obtained constraints on the WD mass and temperature are presented in Figure~\ref{fig:mass-temp}. 

The results show that the companion can indeed be a helium-core WD, as the derived mass lies within the mass range of the model set by \citet{althaus2013}. This mass is compatible with the lower limit $M_c > 0.19$~\msun\ provided by the radio timing measurements (Table~\ref{tab:param}). The derived WD mass and the mass function 0.0026721258~\msun\ obtained by \citet{sanpaarsaphd} yield a lower limit on the pulsar mass of $\ga 1.6$~\msun\  assuming a `median'  orbit inclination  of 60\degs. In a specific case of 90\degs, the lower limit is $\ga 2$~\msun. 
This is in agreement with the fact that 
in MSP-WD binaries neutron stars are on average more massive 
than in double pulsar systems, where the mass 
distribution of neutron stars shows a narrow peak around 
1.35~\msun\ \citep[e.g.,][]{linares}.

%%%%%%%%%%%%%%%%%%%%%%%%%%%%%%%%%%%%%%%%%%%%%%%%%%%%%%%%%%%%%%%%%%%%%%%%%%%
%%%%%%%%%%%%%%%%%%%%%%%%%%%%%%%%%%%%%%%%%%%%%%%%%%%%%%%%%%%%%%%%%%%%%%%%%%%
%%%%%%%%%%%%%%%%%%%%%%%%%%%%%%%%%%%%%%%%%%%%%%%%%%%%%%%%%%%%%%%%%%%%%%%%%%%
\section{Summary and conclusions}
\label{sec:discussion}

Using the GTC, we have performed optical observations of four binary MSPs.
We have detected likely companions to PSRs \psrb, \psrc\ and \psrd\ and 
set the upper limit on the PSR \psra\ companion brightness in the $r'$ band, which is by $\sim$3.6  magnitude deeper than the previous one \citep{scholz2015}. 
The magnitudes and colours of the detected optical sources are consistent   
with the evolutionary sequences of low-mass WDs, confirming the results from the radio-timing measurements. 
Using the WD cooling sequences, we have constrained the parameters of the detected WD companions, 
including mass, temperature and age. The latter two are presented in Table~\ref{tab:red}.

Colours of the PSR \psrb\ companion suggest a WD with a mixed He/H atmosphere. 
The companion has a temperature of about $4\times10^3$ K and age of $\sim2$--5 Gyr.

The PSR \psrd\ companion is consistent with an old ($\ga 5$ Gyr) helium-core hydrogen-atmosphere WD with a mass of 
$0.30^{+0.07}_{-0.06}$\msun\ and a temperature of 4.49(14)~$\times10^3$ K.
Assuming a median orbit inclination of 60\degs, we estimate the minimum pulsar mass to be 1.6 \msun.    

For the PSR \psrc\ system, the optical data 
and the WD cooling predictions suggest that 
the temperature of the companion is about (6--7)$\times10^3$ K regardless of the distance, and its age is $\sim$1--2~Gyr. 
The latest parallactic and the $D_{\rm YMW16}$ distance estimations imply that the WD in the PSR \psrc\ binary may have 
a mixed H/He atmosphere.

Mixed atmospheres in WDs are generally unusual since, due to gravitational settling, pure hydrogen atmospheres are expected \citep[e.g.][]{althaus2000}. Indeed, 
most of the known WD companions to MSPs are known to have hydrogen atmospheres. 
There are, however, several exceptions: the likely ultra-cool companions to PSRs J0751+1807 \citep{bassa2006},
J0740+6620 \citep{beronya2019} 
and  
J2017+0603 \citep{bassa2016}, which may have pure helium or mixed atmospheres; 
the massive CO-core WD companion to the mildly recycled PSR B0655+64 \citep{vankerkwijk1995}, which shows 
carbon lines in its spectrum. Based on the current data, it is possible that the companions to PSRs \psrb\ and \psrc\ can  
complement this set. 
In MSP binaries, WD hydrogen envelopes can be reduced due to irradiation by the pulsar wind following the cessation of the mass transfer \citep{ergma}. Indeed, the above mentioned systems are close binaries ($P_b \la 5$~d) with a high pulsar spin-down power ($\dot{E} \sim 10^{34}$~ erg s$^{-1}$), and it is clear that they can follow this scenario. 
However, since PSRs \psrb\ and \psrc\ possess longer orbital periods, this explanation is unlikely.
To verify this we have obtained basic estimations on the companion flux and the flux of the pulsar wind irradiating 
the companion (see, e.g., \citet{bassa2016}) for both systems. Indeed, the estimations yield a 1--2 orders of magnitude smaller irradiation fluxes in comparison with the companion ones   
implying that in case of PSRs \psrb\ and \psrc\ the irradiation could not have significantly altered the atmospheres of the companions.

Another possibility is that the WD companions to PSRs \psrb\ and \psrc\ have changed their atmospheric compositions from the hydrogen-rich to the helium-rich and back as they cooled down through the convective mixing stage  \citep[e.g.][]{chen2012}:
when the temperature decreases, the surface convection zone of hydrogen expands 
and can reach the underlying helium layer; then the convection brings helium to the surface.
Moreover, other mechanisms changing a WD surface composition were proposed \citep[see e.g.][and references therein]{blouin2019}.

Finally, we find that the WD ages in the PSRs \psra, \psrc\ and \psrd\ binaries are consistent with the characteristic ages of their pulsar companions (see Table~\ref{tab:param}). 
In contrast, the PSR \psrb\ intrinsic characteristic age $\tau_i$, corrected for the Shklovskii and Galactic potential
effects, is 6.1$\pm$0.6 Gyr \citep{sanpaarsaphd}, 
which is somewhat larger than the estimated cooling age of the WD companion. 
However, characteristic ages are rough estimations and 
can significantly deviate form the true pulsar ages \citep[see, e.g.,][]{psrhandbook}. In addition, \citet{tauris2012}
has shown that during the decoupling phase of the companion from its Roche lobe, 
rotational energy of the pulsar is dissipated. As a result, 
characteristic ages do not represent the true age of these
neutron stars. For this reason, in case of PSR \psrb,  
we do not consider the WD and pulsar age inconsistency as a strong argument against the optical identification of the pulsar companion and present the WD cooling ages as independent estimations on the age of the binaries.

As the companions to PSRs \psra, \psrb\ and \psrd\ are very cool, their future studies would mostly rely on broadband near infra-red 
observations where hydrogen and helium WD atmospheres can be best resolved based on the spectral energy distribution \citep{blouin2019}. 
For the warmer PSR \psrc\ companion, optical/near infra-red spectroscopy with 
large-aperture ground-based 
telescopes or with the James Webb Space Telescope might be feasible  
 to get new information on this interesting system, whose   
  pulsar appears to have   an unusually    low mass.

\section*{Acknowledgements}

We thank the anonymous referee for the useful comments that allowed us to improve the manuscript.
We also thank P. Bergeron for providing cooling tracks for WDs with mixed atmospheres.
The work is based on observations made with the Gran Telescopio Canarias (GTC), 
installed at the Spanish Observatorio del Roque de los Muchachos 
of the Instituto de Astrof\'isica de Canarias, in the island of La Palma. 
{\sc iraf} is distributed by the National Optical Astronomy Observatory, 
which is operated by the Association of Universities for Research in Astronomy (AURA) 
under a cooperative agreement with the National Science Foundation. 
This work has made use of data from the European Space Agency (ESA) mission {\it Gaia} 
(\url{https://www.cosmos.esa.int/gaia}), processed by the {\it Gaia} Data Processing and Analysis Consortium 
(DPAC; \url{https://www.cosmos.esa.int/web/gaia/dpac/consortium}).  Funding for the DPAC has been provided by national institutions, 
in particular the institutions participating in the {\it Gaia} Multilateral Agreement. This work also used public data from the North American Nanohertz Observatory for Gravitional Waves (NANOGrav; \url{http://nanograv.org}), which is designated as a ``Physics Frontiers Center" by the National Science Foundation (award \# 1430284).
DAZ thanks Pirinem School of Theoretical Physics for hospitality. The work of AYuK, AVK and DAZ was supported by the Russian Foundation for Basic Research,
project No. 18-32-20170 mol\_a\_ved. The work of SVZ was supported by PAPIIT grants IN-100617 and IN-102120.
 
%%%%%%%%%%%%%%%%%%%%%%%%%%%%%%%%%%%%%%%%%%%%%%%%%%

%%%%%%%%%%%%%%%%%%%% REFERENCES %%%%%%%%%%%%%%%%%%

\bibliographystyle{mnras}

\bibliography{msp}

\bsp	% typesetting comment
\label{lastpage}
\end{document}